# Dark-Field X-Ray Imaging Significantly Improves Deep-Learning based Detection of Synthetic Early-Stage Lung Tumors in Preclinical Models

Joyoni Dey[1], Hunter C. Meyer[1], Murtuza S. Taqi[1]

[1]Department of Physics and Astronomy, Louisiana State University, Baton Rouge, LA, 70808

## Abstract

**Background:** Low-dose computed tomography (LDCT) is the current standard for lung cancer screening, yet its adoption and accessibility remain limited. Many regions lack LDCT infrastructure, and even among those screened, early-stage cancer detection often yield false positives, as shown in the National Lung Screening Trial (NLST) with a sensitivity of 93.8% and a specificity of 73.4% (or a false-positive rate of 26.6%)
**Purpose:** To investigate whether X-ray dark-field imaging (DFI) radiograph—a technique sensitive to small-angle scatter from alveolar microstructure and less susceptible to organ shadowing —can significantly improve early-stage lung tumor detection when coupled with deep-learning segmentation.
**Methods:** Using paired attenuation (ATTN) and DFI radiograph images of euthanized mouse lungs, we generated realistic synthetic tumors with irregular boundaries and intensity profiles consistent with physical lung contrast. A U-Net segmentation network was trained on small patches using either ATTN, DFI, or combined ATTN + DFI channels.
**Results:** The DFI-only model achieved a true-positive detection rate of **83.7%**, compared with **51%** for ATTN-only, while maintaining comparable specificity (90.5% vs. 92.9%). The combined ATTN + DFI input achieved 79.6% sensitivity and 97.6% specificity.
**Conclusion:** DFI substantially improves early-tumor detectability in comparison to standard attenuation radiography and shows potential as an accessible, low-dose alternative for pre-clinical or limited-resource screening where LDCT is unavailable.

## 1. Introduction

Early detection of lung cancer remains one of the most critical determinants of survival, yet the global implementation of low-dose computed tomography (LDCT) screening is far from uniform. For example, a recent study found LDCT utilization rate of only 18.4% among eligible subjects [1]. Another earlier study found screening uptake <6% [2]. Although LDCT is the clinical standard of care, its adoption remains low due to high infrastructure costs, access to rural populations, so on [1-6]. Moreover, the specificity is lower compared to chest radiography resulting in unnecessary follow-up visits and biopsies. For example, the sensitivity and specificity were 93.8% and 73.4% for low-dose CT and 73.5% and 91.3% for chest radiography, respectively [7].
Conventional chest radiography remains the most widely available imaging modality but is inherently limited in detecting small or low-contrast pulmonary nodules due to the overlapping structures of ribs, heart, and mediastinum. X-ray dark-field imaging (DFI) provides a fundamentally different contrast mechanism by detecting small-angle scattering from the alveolar microstructure. This enables visualization of subtle tissue-density variations and pathological changes that are invisible in attenuation-based images.

Here, we present a pre-clinical deep-learning framework combining attenuation and dark-field imaging to improve detection of early-stage lung tumors. Using experimentally acquired mouse dark-field and attenuation radiographs, we generated realistic synthetic tumors spanning 0.75–1.5 mm and trained a patch-based U-Net model to segment tumor regions. We compare single-channel (ATTN-only and DFI-only) and dual-channel (ATTN + DFI) networks to quantify the diagnostic value of DFI. The results demonstrate that DFI dramatically enhances sensitivity without increasing false positives, supporting its potential for affordable and early lung-cancer screening in settings where LDCT is impractical.

## 2. Method

The methods and algorithms are described below.

*Imaging* The imaging experiments were performed by Talbot-Lau X-ray interferometry (TLXI) system at the Pennington Biomedical Research Center, Louisiana State University. The set up and details are given in our prior work [8]. 7 euthanized C57BL/6J (WT) mouse were imaged using the TLXI. All animal-related procedures were approved by the Pennington Biomedical Research Center Institutional Animal Care and Use Committee (IACUC) and were carried out in strict adherence to the guidelines and regulations set by the NIH Office of Laboratory Animal Welfare. The mouse was euthanized via $CO_2$ inhalation, transported to the imaging lab, mounted, and imaged by TLXI.

*Lung Segmentation in Dark-field (DFI)* We segment the lungs on the dark-field (DFI) image and use the attenuation image only as a side-by-side reference during drawing, as shown in Figure 1. Since the lung is the major source of small-angle scatter, the entire lung parenchyma is easier to visualize in the DFI images rather than the attenuation (ATTN) image, in which regions of the lungs occluded by ribs, spine, and the heart.  As seen in Fig. 1(b), although the lower lungs appear completely covered by the other organs in ATTN, they remain visible in DFI Fig 1(a) as low-intensity but detectable regions. Therefore, the DFI image is used as the primary imaging domain for contouring, yielding more anatomically complete lung masks, while the attenuation image is displayed side-by-side to assist contouring in regions where lung boundaries are clearly visible.

Before annotation, the moiré artifacts are removed based on our prior work [8], both ATTN and DFI images are clipped to the [0.1, 99.9] percentile range to make outlines easier to see. Example contours are shown in Figure 1(b), which are filled to create robust L/R lung ROIs that are identical across both the DFI and ATTN channels.

The tumor-insertion step described next uses these masks to place multiple non-overlapping synthetic lesions with approximately spherical projection profiles and irregular boundaries and noise, in both lungs.

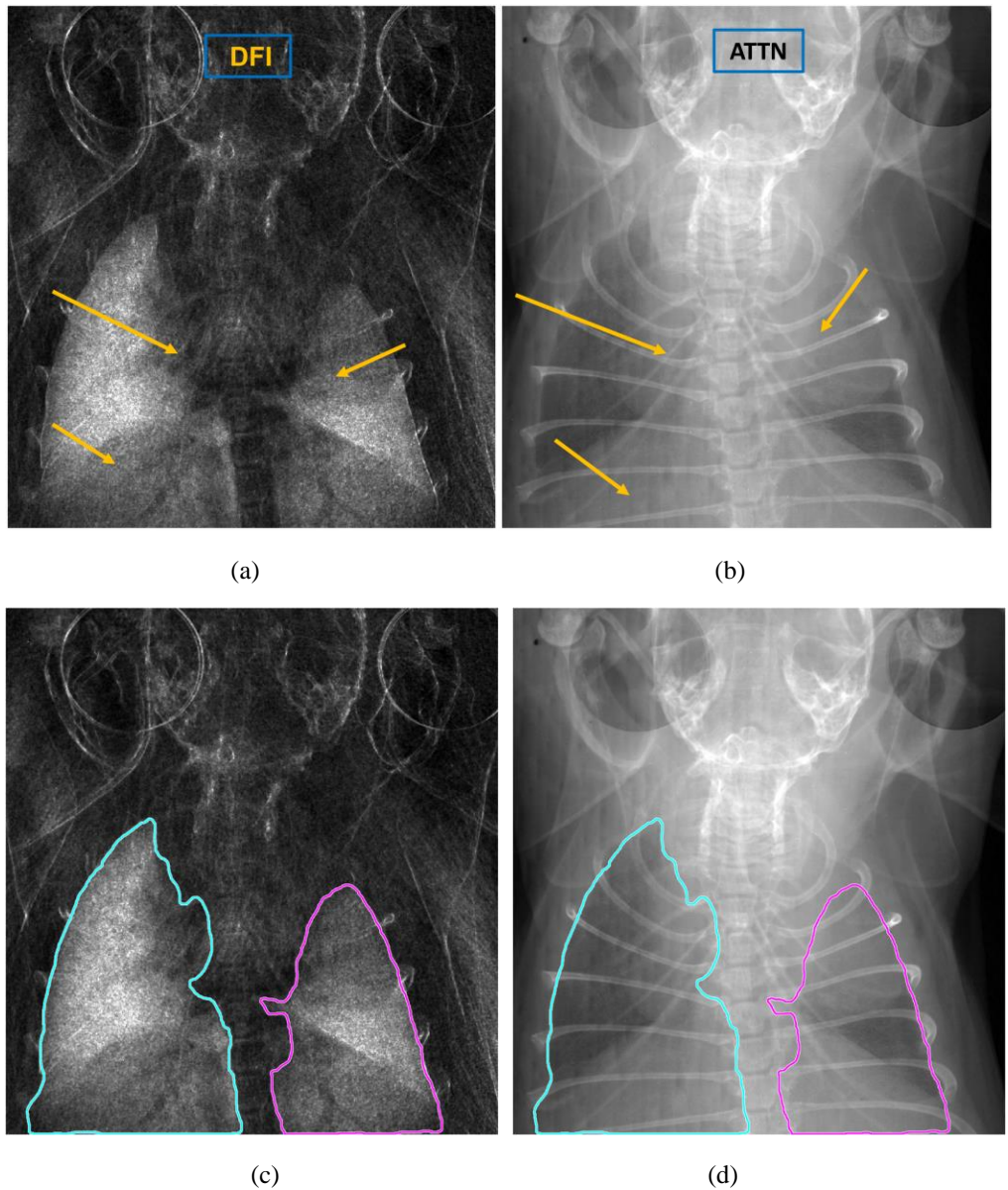

Figure 1. (a) Darkfield (DFI) image (b) Attenuation (ATTN) image. In the attenuation image, several regions of the lungs are partially obscured by the cardiac volume and other organs, whereas in the DFI image these areas remain visible as regions of reduce, but still detectable small-angle scattering intensity. In particular the ribs or cardiac regions emit little or no scatter compared to the porous lungs. (c) and (d) show the contouring performed on the DFI image, with the ATTN image on the side to assist in regions where the lungs are clearly visible in it.

*Realistic tumor insertion* Multiple non-overlapping synthetic tumors were inserted into both the left and right lung regions of mouse images (segmented by the above step in the dark-field images), based on their respective lung masks. Tumors were placed in identical locations in attenuation and dark-field images to maintain channel correspondence.

The central intensity of each tumor is adjusted for attenuation and darkfield, following expected behavior of attenuation and dark-field contrast. Lung tumors have higher attenuation than surrounding lung tissue and appear *brighter* in the standard X-ray attenuation image. This was implemented by an *enhancement factor* (1.05) of the mean upper-lung attenuation, explicitly avoiding lower lung area, to avoid bias of overlapping heart. DFI tumors on the other hand are *darker than the lung dark-field scattering*, so the intensity of the tumor is reduced by a random fraction chosen from **(**0.5 to 0.9**)** of the local DFI mean, consistent with reduced small-angle scattering inside denser tumor regions.

Then, each tumor was applied gray scale shading in intensity by as a *modified projection of a sphere* —whose thickness profile follows $t(r) = \sqrt{1 - \frac{r^2}{R^2}}$ where $R$ = nominal radius of the tumor, and a *parametric gamma* function to further modify the tumor morphology to control the steepness or flatness of the dome. A profile-gamma of 1.5 makes the center steeper for attenuation. A complementary exponent was used for DFI, so that higher the profile-gamma, the center *reduction* is more pronounced. Profile-gamma parameters <1 would make the tumors flatter in ATTN or DFI.

Each tumor boundary is randomized with a *±25% radial jitter* generated using smoothed Gaussian noise over 64 angular samples (can be varied). This avoids perfect circular edges and creates *irregular, realistic tumor perimeters* consistent with biological variability.

To prevent sharp artificial borders, the *transition from tumor to lung was weighted by a smooth step function*. Provisions are also there for local signal-to-noise-matched Gaussian noise with tunable spatial correlation.

A *tumor density parameter* controls how many synthetic lesions are inserted per lung, scaled to its total pixel area. This ensures that smaller lungs contain proportionally fewer tumors and larger ones slightly more, maintaining visual realism.

Figures 2-4 provides representative examples of tumor placement. The original darkfield is shown for reference and to show some inherent dark spots in the dfi image for some mice (Fig. 2(c). All the cases maybe visualized by running *Tumor_insertion.m* .

The tumor density value can be tuned so that when making patches (description follows), the *number of positive (tumor-containing) and negative (tumor-free) patches are roughly balanced*, allowing the subsequent deep-learning classifier to train on comparable datasets without class imbalance.

<u>*Making patches*</u>

Patches (32 × 32 pixels) were extracted from the augmented attenuation and dark-field images within the lung regions to generate paired training inputs for the neural network training and testing. Each patch contained two channels (ATTN and DFI). A neutral network could be fed a single channel or both. Each patch could contain no tumor or one or more tumors. The tumors could be fully enclosed or partially intersecting the patch boundary. A patch is labeled as *positive* if it included *any* tumor pixels, even partially, or *negative* if it contained none. Only patches with at least 50% lung coverage were retained to exclude background and mediastinal regions. Intensity normalization has per-patch or per-mouse options using robust percentile scaling within the lungs to maintain consistent contrast. In our case per-patch normalization was performed.

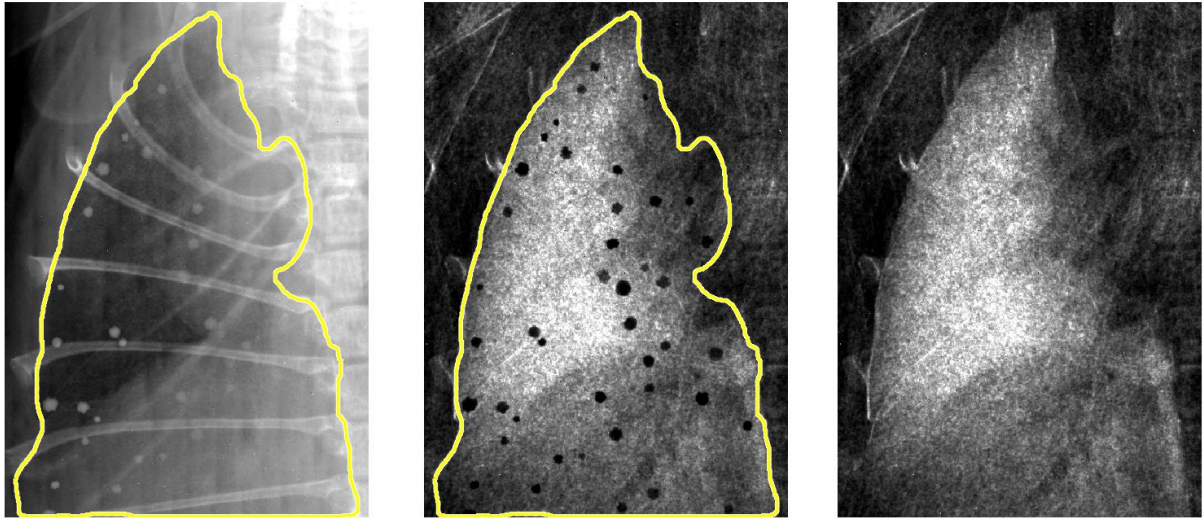

Figure 2. (a) Contoured attenuation (ATTN) image with inserted tumors. The tumors exhibit slightly higher attenuation than the surrounding lung tissue, making them visible in the upper lung regions but nearly indistinguishable in the lower sections, where overlapping organs project higher attenuation. (b) Corresponding contoured dark-field (DFI) image with tumors, where the lesions appear as localized reductions in small-angle scattering. (c) Original dark-field image shown for reference.

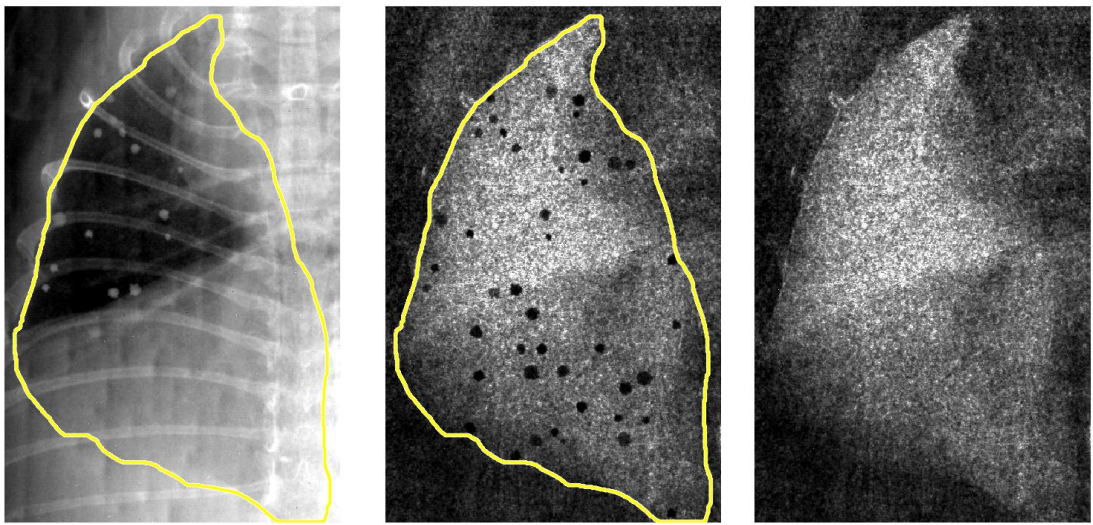

Figure 3. (a) Contoured attenuation (ATTN) image with inserted tumors. The tumors exhibit slightly higher attenuation than the surrounding lung tissue, making them visible in the upper lung regions but nearly indistinguishable in the lower sections, where overlapping organs project higher attenuation. (b) Corresponding contoured dark-field (DFI) image with tumors, where the lesions appear as localized reductions in small-angle scattering. (c) Original dark-field image shown for reference.

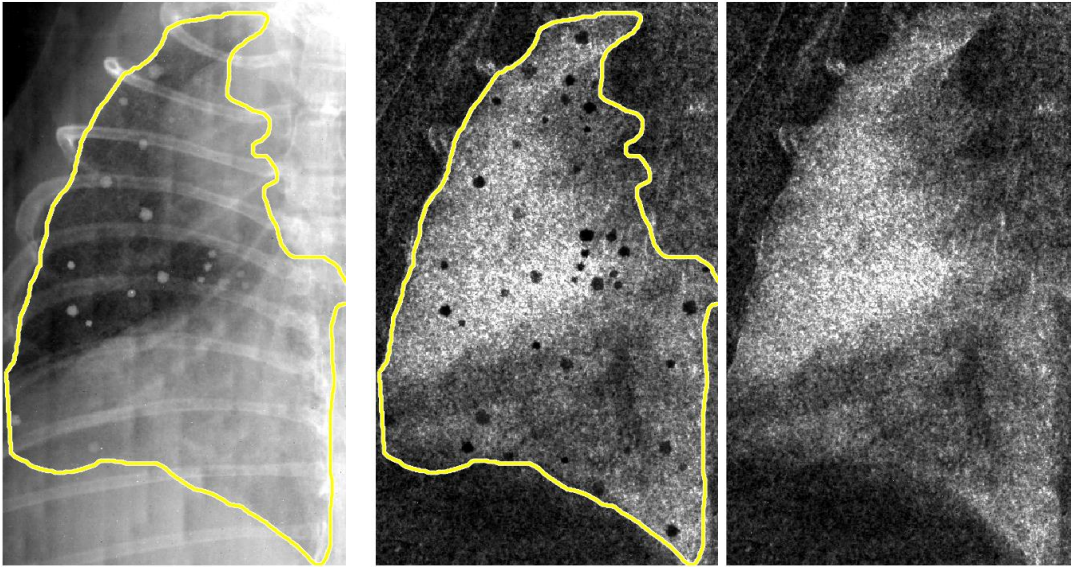

Figure 4. (a) Contoured attenuation (ATTN) image with inserted tumors. The tumors exhibit slightly higher attenuation than the surrounding lung tissue, making them visible in the upper lung regions but nearly indistinguishable in the lower sections, where overlapping organs project higher attenuation. (b) Corresponding contoured dark-field (DFI) image with tumors, where the lesions appear as localized reductions in small-angle scattering. (c) Original dark-field image shown for reference.

### UNET architecture Training/Testing

*2-channel UNET (ATTN and DFI):* A two-dimensional U-Net architecture was used for patch-based tumor segmentation using dual-channel input comprising attenuation (ATTN) and dark-field (DFI) images. Each 32×32 patch for ATTN and corresponding patch for DFI was fed as a 3D tensor of dimension [H x W x C], where H=W=32 and C=2, to jointly exploit the contrast from both imaging modalities. The encoder consisted of two levels, each containing two 3×3 convolutional layers with ReLU activations followed by 2×2 max pooling. The number of feature channels doubled at each down-sampling stage (16–32–64). The decoder mirrored the encoder using 2×2 transposed convolutions for up-sampling and skip connections from corresponding encoder layers to preserve spatial detail. A 1×1 convolution generated two output logits (background and tumor), which were converted to foreground probabilities via a sigmoid or softmax activation.

Training loss was a hybrid binary cross-entropy + Dice loss, Adam optimizer (learning rate = $5\times10^{-4}$, $\beta_1 = 0.9$, $\beta_2 = 0.999$), global-norm gradient clipping (1.0), and early stopping based on validation loss. Random flips and 90° rotations were used for data augmentation.

*1-channel UNET (DFI or Attn):* Two single-input variant of the same U-Net architecture were trained using only the dark-field (DFI) image or the attenuation (ATTN) channel as input. The network preserved the same two-level encoder–decoder structure (16–32–64 feature channels) and output configuration but operated on single-channel 32×32 patches. This design tested the independent discriminative power of DFI contrast or ATTN contrast for tumor detection, isolating its contribution relative to combined **ATTN + DFI** inputs. Training and optimization parameters, data augmentations, and loss functions were identical to the two-channel model for consistent comparison. The testing and validation spits were identical as well for better comparison.

## 3. Results

There were **446** total patches with tumors, some with multiple tumors) and **453** with no tumors. These were split at random 80%-10%-10% into 719 training, 89 validation and 91 testing sets. These same sets were used for training, validation and testing of the three architectures.

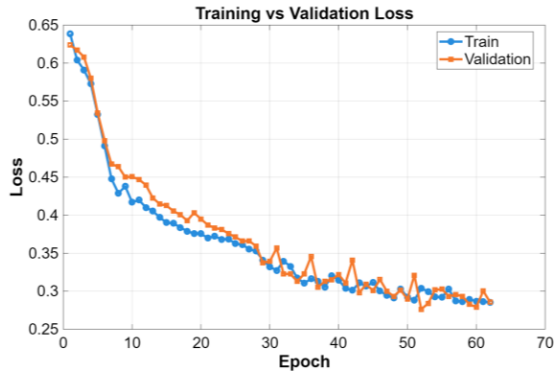 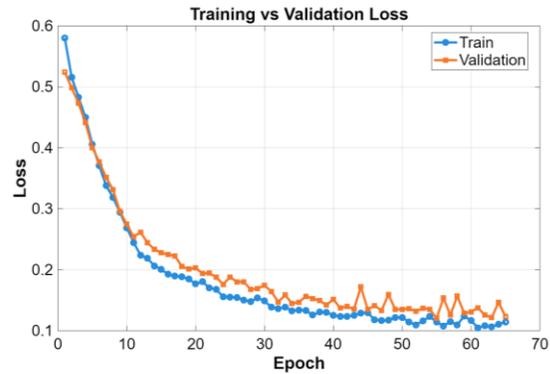

**Figure 5.** For the single-channel **ATTN-only** UNET model, the training and validation losses are shown across epochs. The maximum epoch count was set to 100, but the training stopped early due to oscillations in the validation loss, possibly indicating the onset of overfitting.

**Figure 6.** For the single-channel **DFI-only** UNET model, the training and validation losses are shown across epochs. The maximum epoch count was set to 100, but training stopped early due to oscillations in the validation loss, possibly indicating the onset of overfitting.

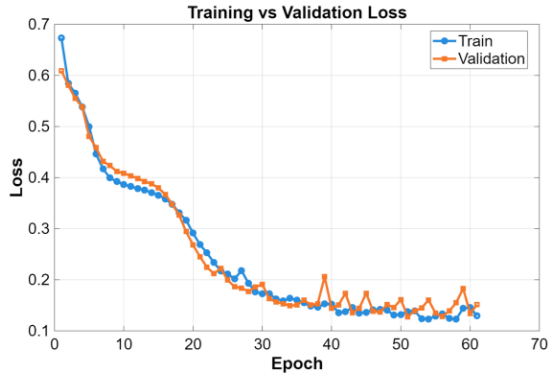

**Figure 7.** For the two-channel **ATTN+DFI** UNET model, the training and validation losses are shown across epochs. The maximum epoch count was set to 100, but the training stopped early due to oscillations in the validation loss, possibly indicating the onset of overfitting.

**Table 1.** Sensitivity (True Positive %) and Specificity (100-False Positive%), both patch-wise, (tumor present/absent in patch). Precision and Recall (pixel-wise classification of each patch).

|  | ATTN-only | DFI-only | ATTN+DFI |
|---|---|---|---|
| **Sensitivity (%TP)** (patch-wise) | 51% | 83.7% | 79.6% |
| **Specificity (100-%FP)** (patch-wise) | 92.9% | 90.5% | 97.6% |
| **Precision** (pixel-wise) | 85.9% | 87.8% | 94.2% |
| **Recall** (pixel-wise) | 44.7% | 85.5% | 78.2% |

The training results of the single-channel UNET with ATTN-only and DFI-only are shown in Fig. 5-6. Fig. 7 shows that of 2-channel architecture of ATTN+DFI. The Table 1 shows that the true-positive sensitivity improved with dark-field only patches to 83.75 from 51% with just attenuation patches. The Specificity was slightly better with attenuation 92.9% (ATTN-only) versus 90.5% with DFI-ONLY. The ATTN+DFI has intermediate results of Sensitivity of 79.6% and improved Specificity to 97.6%.

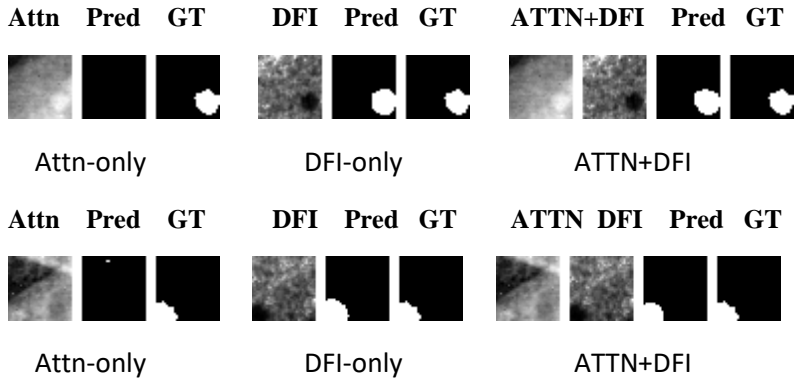

**Figure 8.** The test panels show the patch(es), predicted mask and the ground truth (GT) masks for Attn-only, DFI-only and ATTN+DFI UNET results. The top and bottom panels show patches where the ATTN-only misses a tumor but DFI-only or ATTN+DFI finds the tumor.

## 4. Discussions and Conclusion

The DFI-only showed tremendous improvement of detection sensitivity compared to attenuation radiographs. The false-positive (1-sensitivity) was comparable if slightly lower for darkfield. The tumor range was 0.75-1.5mm in this project.

If these results translate to the clinic, attenuation and dark-field radiography along with a AI interpreter may well full-fill a gap in clinics where Low-Dose-CT is not accessible or for patients are not covered.

The dark-field performance can be potentially further improved with denoising. More data will allow larger range of tumor sizes and intensities both in the dark-field and attenuation.

## 5. Acknowledgments

We wish to thank Dr. Leslie Butler and Dr. Kyungmin Ham for their help in upgrading and automating the Keck System at Pennington Biomedical Research Center. We wish to thank Dr. Christopher Morrison for providing the deceased mice that was imaged at the Keck System and used for this project.

**Author Contributions**: JD is responsible for designing the tumor inserts and AI architecture. HCM and MST and JD acquired the mice data.

**Data Availability:** The data, code and instructions for this project can be obtained at https://github.com/deyj/dfi-atn_unet-lung-tumor_det/ or asking the first author.